\documentclass[aps,pre,reprint,superscriptaddress]{revtex4-1}
\usepackage{graphicx}
\usepackage{amssymb}
\usepackage{amsmath}
\usepackage{hyperref}
\usepackage{float}
\usepackage{mathtools}

\begin{document}

\title{Attainment of Herd Immunity: Mathematical Modelling of Survival Rate}

\author{Sayantan Mondal}
\affiliation{Solid State and Structural Chemisry Unit, Indian Institute of Science, Bengaluru-560012, India}
\author{Saumyak Mukherjee}
\affiliation{Solid State and Structural Chemisry Unit, Indian Institute of Science, Bengaluru-560012, India}
\author{Biman Bagchi}
\affiliation{Solid State and Structural Chemisry Unit, Indian Institute of Science, Bengaluru-560012, India}

\begin{abstract}
We study the influence of the rate of the attainment of herd immunity (HI), in the absence of an approved vaccine, on the vulnerable population. We essentially ask the question: \textit{how hard the evolution towards the desired herd immunity could be on the life of the vulnerables}? We employ mathematical modelling (chemical network theory) and cellular automata based computer simulations to study the human cost of an epidemic spread and an effective strategy to introduce HI. Implementation of different strategies to counter the spread of the disease requires a certain degree of quantitative understanding of the time dependence of the outcome. In this paper, our main objective is to gather understanding of the dependence of outcome on the rate of progress of HI. We generalize the celebrated SIR model (Susceptible-Infected-Removed) by compartmentalizing the susceptible population into two categories- (i) vulnerables and (ii) resilients, and study dynamical evolution of the disease progression. \textit{We achieve such a classification by employing different rates of recovery of vulnerables vis-a-vis resilients}. We obtain the relative fatality of these two sub-categories as a function of the percentages of the vulnerable and resilient population, and the complex dependence on the rate of attainment of herd immunity. Our results quantify the adverse effects on the recovery rates of vulnerables in the course of attaining the herd immunity. \textit{We find the important result that a slower attainment of the HI is relatively less fatal}. However, a slower progress towards HI could be complicated by many intervening factors.
\end{abstract}

\maketitle

\section{Introducntion}

The present COVID-19 pandemic is a dynamic and volatile process with often unpredictable ups and downs in the infected populations that make it difficult to predict its future course. In the absence of any vaccine or definitive drug in the immediate future \cite{Chen2020} the fight against COVID-19 is a hard and long drawn bitter battle, with two strategies being put forward. The first is the widely enforced lockdown, quarantine, and social distancing where the spread of the disease is contained at its inception and only a limited fraction of population is allowed to be infected.\cite{Prem2020} This model appears to be successful in South Korea and China, and some other Asian countries.\cite{Shen2020} The other model is to allow the virus to have a relatively unconstrained transmission so that a large fraction of the people develops the immunity.\cite{Kamikubo2020} This is called the herd immunity (HI) that is favoured by Sweden, and was initially discussed by Germany and England, but largely discarded later. HI can be achieved by two ways- (i) by vaccination, and (ii) by infection. The HI approach is based on the understanding that one can obtain the herd immunity in the society if 60-70\% of the population gets immunized. Needless to say this herd immunity is preferable through vaccination as happened in small pox and measles. Implementation of both the models has difficulties. Implementation of lockdown and social distancing requires enormous effort, backed up by resources. On the other hand, the HI model could have adverse consequence on the vulnerable citizens, a subject not adequately discussed. In fact, experiences in Italy and Spain show that the demography can be altered in some regions if HI is given an unconstrained run.

\begin{figure}[H]
 \centering
 \includegraphics[width=3.2in,keepaspectratio=true]{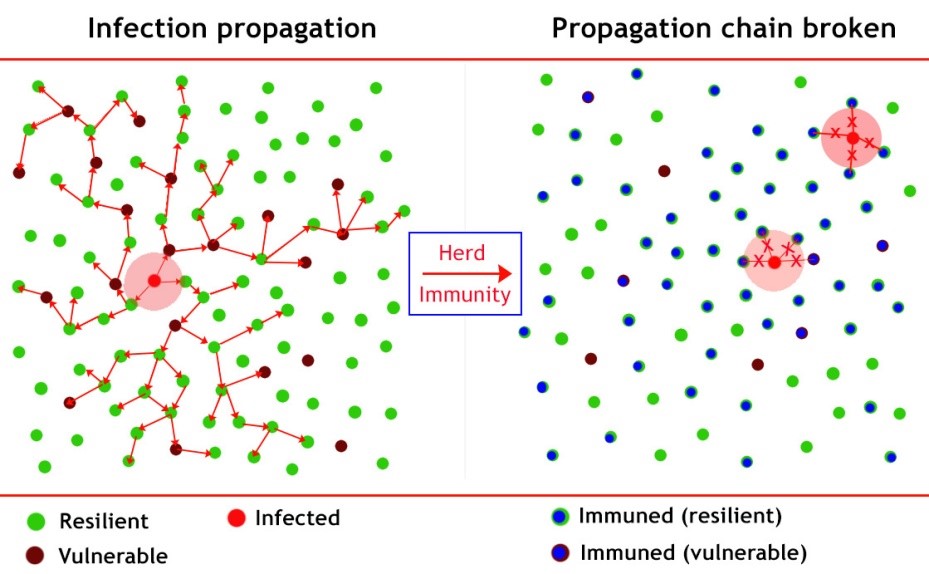}
 \caption{A pictorial representation of the herd immunity phenomenon. In the left we have a region with the susceptible population and one infected person. The total susceptibles are further divided into vulnerables and resilients. The infection propagates in an unconstrained manner and after a certain period the region possesses a large fraction of immunized population (right). After this immunisation any further infection cannot propagate and indirectly protects the susceptibles. In addition to that, multiple infected persons cannot do further harm. The colour codes are maintained throughout this paper.}
 \label{fig1}
\end{figure}

Herd immunity ensures an indirect protection from COVID-19 (or any other infectious disease) when a large fraction of the population becomes immunized to the disease.\cite{Fine1993,Anderson1985,John2000} Herd immunity drastically decreases the probability of the presence of an uninfected individual in the vicinity of a presently infected individual. That is, the infected person is effectively quarantined by the surrounding immunized population. Hence, the chain of propagation breaks. In Fig. \ref{fig1} we pictorially explain the phenomenon of herd immunity.

This can happen by providing the population with a vaccine or by getting cured after being infected. In the case of COVID-19 pandemic, as of now, we are unsure regarding the success of a vaccine and the latter is the only option to attain HI. However, the herd immunity threshold (HIT), that is the minimum fraction of population needs to get immunized in order to eradicate the disease, is different for different infectious diseases.\cite{Georgette2009,McBryde2009} For example, HIT for measles is ~92-95\% and for that of SARS-1 it is in the range of 50-80\%.

Researchers around the world are exploring mainly two aspects of this disease- (i) the microscopic and clinical aspects which would eventually lead to drug discovery and vaccine preparation,\cite{Chen2020,Wrapp2020} (ii) the demographic aspects which lead to policy making and timeline prediction.\cite{Prem2020,Shen2020,Singh2020,Mukherjee2020} The latter requires effective mathematical modelling and crowd simulations. However, these models fail to predict the real scenario because of some inherent assumptions and limitations. Although a lot of interesting new studies are emerging in both categories in the context of the recent coronavirus pandemic, the issue of herd immunity and its fatality are not studied.

There are several mathematical models which have been employed in the context of epidemic modelling, for example, the famous Kermack-McKendrick (KM) model which has been used extensively to study the spread of infectious diseases like measles, small pox etc.\cite{Daley2001,Kermack1927} At the core of this model lies a system of three coupled differential equations for susceptible (S), infected (I) and removed (R) (cured and dead) populations, that is, the famous SIR model (Eq. \ref{eq1}).\cite{Skvortsov2007,Jones2009,Anderson1979} At the onset of an epidemic S becomes I and I eventually becomes R, but R can never become S or I because of acquired immunity.  

\begin{equation}
 \begin{split}
  \frac{dS}{dt}& =-k_{S\rightarrow I}SI\\
  \frac{dI}{dt}& =k_{S\rightarrow I}SI-k_{I\rightarrow R}I\\
  \frac{dR}{dt}& =k_{I\rightarrow R}I
 \end{split}
 \label{eq1}
\end{equation}

Eq. \ref{eq1} describes the three coupled non-linear differential equations of the KM model where $k_{S\rightarrow I}$ is the rate of infection and $k_{I\rightarrow R}$ is the rate of removal (recovery and death). In the conventional SIR model $k_{S\rightarrow I}$ and $k_{I\rightarrow R}$ are written as $\alpha$ and $\beta$ respectively. In principle the rate constants should be time and space dependent, that is, non-local in nature. But it is difficult to predict the functional form of the rate constants with time- it could be periodic, decaying or stochastic in nature. The applicability of this model is for a homogeneous population distribution and mass transmission at a large scale.\cite{Daley2001}

An important quantity is the basic reproduction number ($R_0$) which is an estimate of the number of secondary infection from one primary infection.\cite{Dietz1993} The value of $R_0$ is intimately connected with the herd immunity threshold ($H_t$) discussed above.\cite{McBryde2009,Diekmann1995} (Eq. \ref{eq2}) Hence a correct determination of the basic reproduction parameter, $R_0$, is important.

\begin{equation}
 H_t=\left(1-\frac{1}{R_0}\right)\times 100 \%
 \label{eq2}
\end{equation}

It is clear from Eq. \ref{eq2} that a higher value of $R_0$ increases the herd immunity threshold. For SARS-Cov2 the value of $R_0$ shows a large dispersion and as a consequence we cannot predict the value of $H_t$. For COVID-19 the average value of $R_0$ is estimated to be in the range of $\sim$2.0-3.0 but it can possess spatial heterogeneity and time dependence in reality.\cite{Zhang2020,Tang2020} If one considers $R_0$ to be in the range of 2.0-3.0 the value of $H_t$ would be in between 50\%-66\%.

In the light of SIR model [Eq.(1)] $R_0$ can be defined as

\begin{equation}
 R_0=\frac{k_{S\rightarrow I}}{k_{I\rightarrow R}} S
 \label{eq3}
\end{equation}

Eq. \ref{eq3} provides a different definition of $R_0$ and can be understood as follows.  If we assume that (the S the fraction of susceptible population) is near 1.0 at the beginning (as there are very few infections compared to a huge population), then $R_0$ could be equal to unity if the two rate constants are equal. This means that the number of infection and recovery are same at any time. In this situation the disease remains under control. $R_0 > 1$ causes an epidemic as it challenges the capacity of the healthcare facilities. However, for different region the value of $R_0$ could be different depending on the intensity of region wise preventive and healthcare measures.

In this work we ask the following questions- (i) what are the relative magnitude of the fatality to the vulnerable and resilient populations if we attempt to achieve HI without a vaccine? (ii) What is the dependence of the fraction of survival on the rate of the attainment of HI? These two issues are widely discussed all over the world. Here we seek answers to these two important questions by employing a modified Susceptible-Infected-Removed (SIR) model and cellular automata (CA) simulations.

The rest of the paper is organised as follows. In section \ref{sec2} we describe the mathematical model and the CA simulation protocols. Section \ref{sec3} consists of the results from numerical solutions of the modified SIR model and simulations, accompanied by detailed discussions. This section is further divided into several sub sections. In section IV we summarize and conclude our study.

\section{Theoretical Formalism}
\label{sec2}

\subsection{Mathematical Modelling}

\begin{figure}[ht]
 \centering
 \includegraphics[width=3in,keepaspectratio=true]{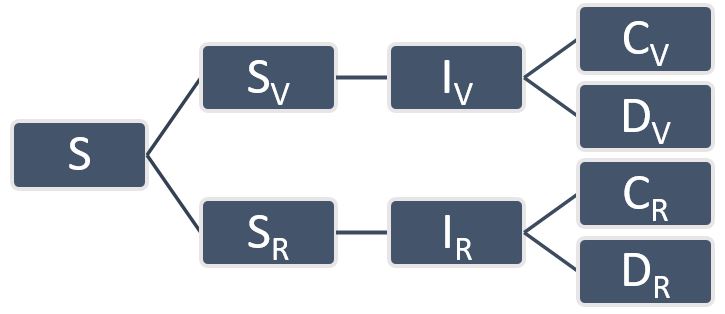}
 \caption{Schematic representation of the modified SIR network model. Here the susceptible (S) population is divided into $S_V$ and $S_R$ that represent elderly and younger people respectively. A part of the fraction $S_V$ gets infected and creates $I_O$ fraction of infected population. A part of the remaining fraction of the population, that is, $S_R$ gets infected and creates $I_R$ fraction of the infected population. Both $I_V$ and $I_R$ get either cured (C) or dead (D). Naturally the rate of recovery for the younger fraction of the population is more than that of the older infected population. On the other hand, the rate of death for the older population is more than that of the younger invectives.}
 \label{fig2}
\end{figure}

We modify the celebrated SIR (Susceptible-Infected-Removed) model by dividing the entire susceptible population into two parts, namely vulnerable (Vul) and resilient (Res). In the context of the corona virus disease, the vulnerable category consists of persons who are above 60 years of age or have pre-existing medical conditions like diabetes, heart and kidney disease, and lung conditions.\cite{Yang2020} The rest of the population is termed as resilient who have a greater chance of getting cured. We achieve such classification by employing different rate constants associated with their recovery. This is based on the available data on the coronavirus disease. The scheme of this classification is described in Fig. \ref{fig2}.

We follow the scheme described in Fig. \ref{fig2} and formulate a system of eight coupled non-linear differential equations [Eqs. \ref{eq4} - \ref{eq11}].

\begin{equation}
 \frac{dS_{Vul}(t)}{dt} = -k_{S_{Vul}\rightarrow I_{Vul}}(t)S_{Vul}(t)I(t)
 \label{eq4}
\end{equation}

\begin{equation}
 \frac{dS_{Res}(t)}{dt} = -k_{S_{Res}\rightarrow I_{Res}}(t)S_{Res}(t)I(t)
 \label{eq5}
\end{equation}

\begin{equation}
\begin{split}
 \frac{dI_{Vul}(t)}{dt} = k_{S_{Vul}\rightarrow I_{Vul}}(t)S_{Vul}(t)I(t)\\ -(k_{I_{Vul}\rightarrow C_{Vul}}(t)+k_{I_{Vul}\rightarrow D_{Vul}}(t))I_{Vul}(t)
 \end{split}
 \label{eq6}
\end{equation}

\begin{equation}
\begin{split}
 \frac{dI_{Res}(t)}{dt} = k_{S_{Res}\rightarrow I_{Res}}(t)S_{Res}(t)I(t)\\ -(k_{I_{Res}\rightarrow C_{Res}}(t)+k_{I_{Res}\rightarrow D_{Res}}(t))I_{Res}(t)
 \end{split}
 \label{eq7}
\end{equation}

\begin{equation}
 \frac{dC_{Vul}(t)}{dt} = k_{I_{Vul}\rightarrow C_{Vul}}(t)I_{Vul}(t)
 \label{eq8}
\end{equation}

\begin{equation}
 \frac{dC_{Res}(t)}{dt} = k_{I_{Res}\rightarrow C_{Res}}(t)I_{Res}(t)
 \label{eq9}
\end{equation}

\begin{equation}
 \frac{dD_{Vul}(t)}{dt} = k_{I_{Vul}\rightarrow D_{Vul}}(t)I_{Vul}(t)
 \label{eq10}
\end{equation}

\begin{equation}
 \frac{dD_{Res}(t)}{dt} = k_{I_{Res}\rightarrow D_{Res}}(t)I_{Res}(t)
 \label{eq11}
\end{equation}

In the following, we explain the complex set of equations. Here I(t) is the number of total infectives at any time t, that is $I(t)=I_{Vul}(t)+I_{Res}(t)$. This is the variable that couples the two population sub-categories. $k(t)$ are the rate constants associated with processes that are described in the subscript with an arrow. 

We would like to point out that the rates in above equations of motion are all assumed to be time dependent. These rate constants contain all the basic information and also connected with $R_0$. In our earlier study, we employed a time dependent rate to produce certain features observed in the time dependence of new cases such a double peaked population structure.\cite{Mukherjee2020} The time dependence of rate can be employed to include certain dynamical features like crossover from local contact to community transmission. It is worth stressing that the modelling of these time dependent rate constants plays a pivotal role in the SIR scheme.

We propagate these equations numerically to obtain the respective temporal evolution of each kind of population fraction. From the temporal profiles we can extract several important quantities after a long time (that is, the end of the spread), for example, (i) the peak height of the active infected cases, (ii) the fraction of cured population, (iii) the fraction of dead population, (iv) the fraction of uninfected population, (v) time required to reach the immunity threshold etc. We can regard these equations together to form a system of reacting species, as in a system of chemical reactions.

We solve these equations with two different sets of the rate constant values and aim to understand the relative damages to the vulnerable and resilient population. The values of rate constants are provided in Table \ref{tab1}. We keep $k_{S_{Vul}\rightarrow I_{Vul}}$ and $k_{S_{Res}\rightarrow I_{Res}}$ the same which depicts the same probability of getting infected for both the sub-categories. However, the rate constants associated with recovery and death differs in orders of magnitude between Vul and Res.

We now discuss the procedure we follow to assign different rate constants to the vulnerables and resilients. In a previous study we estimated the values of $k_{S\rightarrow I}$ and $k_{I\rightarrow R}$ by fitting the infected/cured/death vs. time data for India (source: www.covid19india.org).\cite{Mukherjee2020} We plot the rate of change of the cured ($dC/dt$) and dead ($dD/dt$) population against the infected population to find the slope that gives the rate. This procedure provides us with required estimates of $k_{I\rightarrow C}$ and $k_{I\rightarrow D}$. For India, till $27^{th}$ May, the estimated values are $k_{I\rightarrow C} = 0.026 \:day^{-1}$ and $k_{I\rightarrow D} = 0.0013 \:day^{-1}$. That is, $k_{I\rightarrow C}$ is approximately 20 times of $k_{I\rightarrow D}$. However, for countries like Italy, Spain, and USA $k_{I\rightarrow D}$ was significantly higher. This comparison however takes no cognition of the relative time scales, and therefore should be taken with care.

These values are mean field in nature and contain enormous spatial heterogeneity. If we see the state wise (or district wise) statistics we find a large dispersion. On the other hand, the determination of $k_{S\rightarrow I}$ is not that straight forward as the equations containing $k_{S\rightarrow I}$ are non-linear in nature in the SIR model. Hence one needs to obtain a good estimate of $R_0$ and calculate $k_{S\rightarrow I}$ from Eq. \ref{eq3}. As mentioned above, $R_0$ also exhibits spatiotemporal heterogeneity which makes the problem of estimating the rate constants even more challenging. For example, in Italy $R_0$ has been estimated to be $\sim$3.0-6.0 and in the Hunan province of China it is $\sim$1.73-5.25.\cite{Wangping2020} In a recent study on Wuhan, the transmission rate ($k_{S\rightarrow I}$) is assumed to vary from 0.59 to $1.68 \:day^{-1}$.\cite{Lin2020}

However, the data required to extract the rate constants associated with the two individual sub-categories, namely, vulnerable and resilient, are not available separately. As the values of the rate constants are connected to the basic reproduction number ($R_0$), we choose the inputs, by preserving the basic features, such that the average value of $R_0$ yields an acceptable number, in light of acquired information. Next we tune the parameters such that the maximum of the active cases falls in the range of $\sim$60-90 days, as observed for most countries. We note that we consider these values only to study the trends and do not strictly correspond to any particular region in reality.

\begin{table}[ht]
\caption{The values of rate constants used to solve the system of coupled differential equations [Eq.\ref{eq4} - \ref{eq11}]. The unit of the rate constants is $day^{-1}$.}
 \centering
 \begin{tabular}{|c|c|c|}
  \hline
  Rate Const. & Set-1 & Set-2\\
  \hline
  \hline
  $k_{S_{Vul}\rightarrow I_{Vul}}$ & 0.50 & 0.78\\
  \hline
  $k_{I_{Vul}\rightarrow C_{Vul}}$ & 0.05 & 0.05\\
  \hline
  $k_{I_{Vul}\rightarrow D_{Vul}}$ & 0.10 & 0.10\\
  \hline
  $k_{S_{Res}\rightarrow I_{Res}}$ & 0.50 & 0.78\\
  \hline
  $k_{I_{Res}\rightarrow C_{Res}}$ & 0.50 & 0.50\\
  \hline
  $k_{I_{Res}\rightarrow D_{Res}}$ & 0.05 & 0.05\\
  \hline
  \end{tabular}
 \label{tab1}
\end{table}

We invoke two different values of $R_0$ for the two different sub-categories. For set-1 $R_0^{Vul} = 3.33$ and $R_0^{Res} = 3.33$. The larger value of $R_0$ for vulnerables arise from slower rate of recovery, Eq. \ref{eq3}.

On the other hand, for set-2 $R_0^{Vul} = 5.20$ and $R_0^{Res} = 1.42$. We obtain these values by considering each of the population to be individually normalised (that is ~100\%). In such a situation the effective $R_0$ can be calculated as follows (Eq. \ref{eq12}).

\begin{equation}
 R_0^{eff} = \frac{R_0^{Vul}N_{Vul}+R_0^{Res}N_{Res}}{N_{Vul}+N_{Res}}
 \label{eq12}
\end{equation}

Here $N_{Vul}$ and $N_{Res}$ represent the number of people in the vulnerable and resilient category respectively. In all our calculations we start with total infected fraction as 0.001 and vary the percentage of vulnerable populations from 5\%-40\%.  By using Eq. \ref{eq12} we calculate the effective $R_0$ values for different ratio of vulnerable to resilient population. We find $R_0$ varies from 1.03 to 1.88 for set-1 and 1.61 to 2.93 for set-2. In a way, set-1 represents a more controlled situation compared to set-2 (Table \ref{tab2}).

\begin{table}[ht]
\caption{The basic reproduction number ($R_0$) for the parameters described in Table \ref{tab1} (set-1 and set-2) for various ratios of vulnerable to resilient population.}
 \centering
 \begin{tabular}{|c|c|c|c|}
  \hline
  \% & \% & \multicolumn{2}{c|}{$R_0$} \\
  \cline{3-4}
  vulnerble & resilient & Set-1 & Set-2 \\
  \hline
  \hline
  5  & 95 & 1.031 & 1.609 \\
  \hline
  10 & 90 & 1.152 & 1.798 \\
  \hline
  15 & 85 & 1.273 & 1.987 \\
  \hline
  20 & 80 & 1.394 & 2.176 \\
  \hline
  25 & 75 & 1.515 & 2.365 \\
  \hline
  30 & 70 & 1.636 & 2.554 \\
  \hline
  35 & 65 & 1.757 & 2.743 \\
  \hline
  40 & 60 & 1.878 & 2.932 \\
  \hline
  \end{tabular}
 \label{tab2}
\end{table}

\subsection{Stochastic Cellular Automata Simulation}

Stochastic cellular automata (CA) simulations give a microscopic and nonlocal picture of the problem at hand. Such simulations are often used to model several physical phenomena.\cite{Hollingsworth2004, Seybold1998, Wolfram1983, Bartolozzi2004, Soares-Filho2002, Goltsev2010, Almeida2011} Unlike the mathematical model, CA simulations can directly establish a physical map of the disease-spread. Moreover, we incorporate several region specific and disease specific parameters in our CA simulations that give a general outlook to our investigations. A detailed list of the parameters and associated symbols can be found in our previous work.\cite{Mukherjee2020}

The spread of COVID-19 is strongly inhomogeneous. So, a homogeneous model fails to capture many aspects. In a real-world scenario, the non-local description may often become important in determining the fate of a pandemic in a given geographical region. In such a case, the population parameters are space-dependent. Moreover, the rate constants also have a spatial distribution. Hence, solutions of these equations are highly non-trivial and a large scale cellular automata simulation may capture these inherent spatiotemporal heterogeneities.

In this work, we neglect the effects of social distancing and quarantine, since we aim at establishing a relation between the percentage of mortality and immunization by an unhindered transmission of the disease within the whole population. Calculation of the rates of transmission and recovery/death can often be difficult due to several reasons like unavailability of data, political or demographic complications etc. This becomes particularly nontrivial when we consider the process with respect to a given population distribution of vulnerable and resilient individuals. The probabilistic approach employed in our simulations makes it easier to study the process, since obtaining an average probability for each of the processes is much more practical.

We use the Moore definition \cite{Fu2003, White2007, Sirakoulis2000} to denote the neighbourhood of a given person.  The salient features of our simulation are detailed in our previous work.\cite{Mukherjee2020} Here, we summarize our CA simulation methodology. 

We start we a land randomly occupied by susceptibles and infectives. The population distribution is such that  5\% and 0.05\% of the total available land is covered by susceptibles and infectives respectively. We divide the population into vulnerable and resilient individuals with respect to their probabilities of recovery ($P_R^{Vul}$ and $P_R^{Res}$). Vulnerables primarily include people above the age of 60. This also includes people with serious health issues, who are more prone to get deceased if infected.\cite{Remuzzi2020, Ruan2020, Wu2020} The resilients, on the other hand, are the young fraction of the society with no severe health conditions. When an infective comes in the neighbourhood of a susceptible, the latter is converted to an infective with a given probability of transmission which is considered to be equal and time independent (constant) for both vulnerables and resilients. The time period of infection is determined by probability of recovery and the probability of remaining infected in a given simulation step. In this work, we consider the latter to be 0.99.\cite{Mukherjee2020} An individual, once cured from infection, becomes immune to the disease.

We run our simulations for a given number of steps ($N$). It should be noted that the time unit is not well-defined for this simulations. To get an estimate of time, the results need to be compared with our theoretical model. 

\section{Results and Discussion}
\label{sec3}
\subsection{Numerical solutions of the SIR model}
\label{sec3A}

\begin{figure}[ht]
 \centering
 \includegraphics[width=3.4in,keepaspectratio=true]{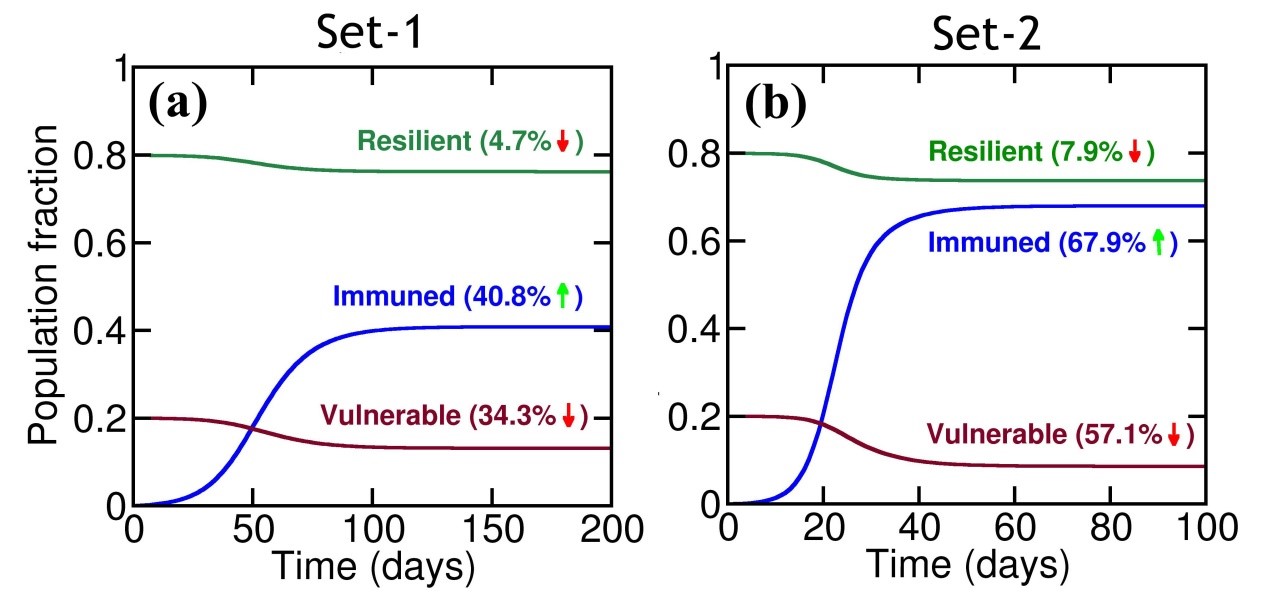}
 \caption{Population disease progression as obtained from the  solution of the system of eight coupled non-linear differential equations presented in Eqs. \ref{eq4} - \ref{eq11} as function of time for two different situations described in Table \ref{tab1}. Plots show the increase in the total immunity (blue) with the decrease in the populations of vulnerable population (maroon) and resilient population (green) for (a) Set-1 and (b) Set-2. In these two calculations we start with $V:R=1:4$. In both the two cases the percentage demise in the vulnerable population is significantly higher.}
 \label{fig3}
\end{figure}

Here we present the results from the numerical solutions of Eqs. \ref{eq4} - \ref{eq11} in Fig. \ref{fig3}. We choose two sets of rate constants, set-1 (Fig. \ref{fig3}a) and set-2 (Fig. \ref{fig3}b) and obtain the changes in the population of vulnerables and resilients. With our choice of parameters (Table \ref{tab1}) for set-1 we observe 40.8\% increase in the immuned population. In order to achieve the 40.8\% immunity a region loses 4.7\% of its resilient population and 34.3\% vulnerable population. On the other hand, for set-2 a region loses 7.9\% of its resilient population and 57.1\% of its vulnerable population in order to achieve $\sim$68\% immunity (that could be the HIT for COVID-19). Hence, it is clear that for both the two cases the vulnerables are significantly affected. We note that with an increased infection rate the timescale of the saturation of the temporal profiles are drastically reduced. The graphs that are presented in Fig. \ref{fig3} are obtained for 20\% initial vulnerable population.

\begin{figure}[h]
 \centering
 \includegraphics[width=3.2in,keepaspectratio=true]{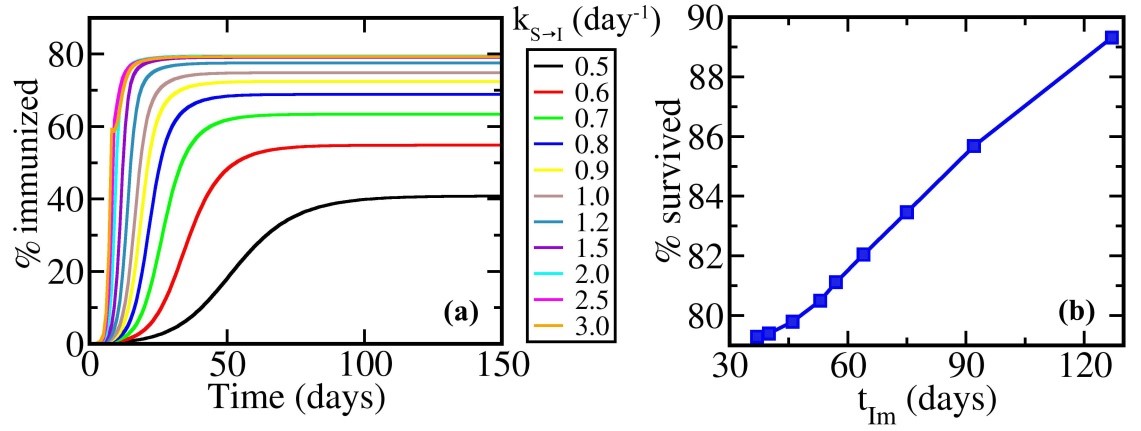}
 \caption{The effect of different rates of attaining herd immunity on the total population. (a) Plot of the time evolution of the percentage of total immunized population for different values of susceptible to infected rate-constants. With increasing $k_{S\rightarrow I}$ we see an increase in the percentage immunity and decrease the time required to reach saturation ($t_{Im}$). (b) Percentage survival (uninfected and cured population) of the total population against $t_{Im}$. The two quantities show linear dependence. That is, the percentage survival increases as we take more time to reach immunity saturation. Note that both the X and the Y axes are the outcome of the numerical solution and not provided as inputs. The calculations are done using a fixed Vul:Res=1:4 and the rate constants associated with recovery/death are also kept same as given in Table \ref{tab1}.}
 \label{fig4}
\end{figure}

In Fig. \ref{fig4}a, we show the time evolution of the total immunity percentage. In order to study the effect of fast (early) vs slow (late) achievement of the immunity saturation, we plot the percentage survival of the total population against the time required to attain the immunity threshold ($t_{Im}$) for different values of $k_{S\rightarrow I}$ (Fig. \ref{fig4}b). We find that the percentage of survival increases linearly with increasing $t_{Im}$. This indicates that a quick achievement of immunity saturation could lead to fatal consequences. \textit{If a society opts for herd immunity, it has to be a slow process}. 

To make the immunity gaining process slow (which leads to relatively less casualty), the rate of infection ($k_{S\rightarrow I}$) needs to be brought down. On the other hand, the rate of removal (recovery and death), $k_{I\rightarrow R}$, depends primarily on the disease and partly on the presently available healthcare facilities. $k_{S\rightarrow I}$ can be controlled by employing effective strategies like lockdown, quarantine, and social distancing.   

\begin{figure}[H]
 \centering
 \includegraphics[width=3.2in,keepaspectratio=true]{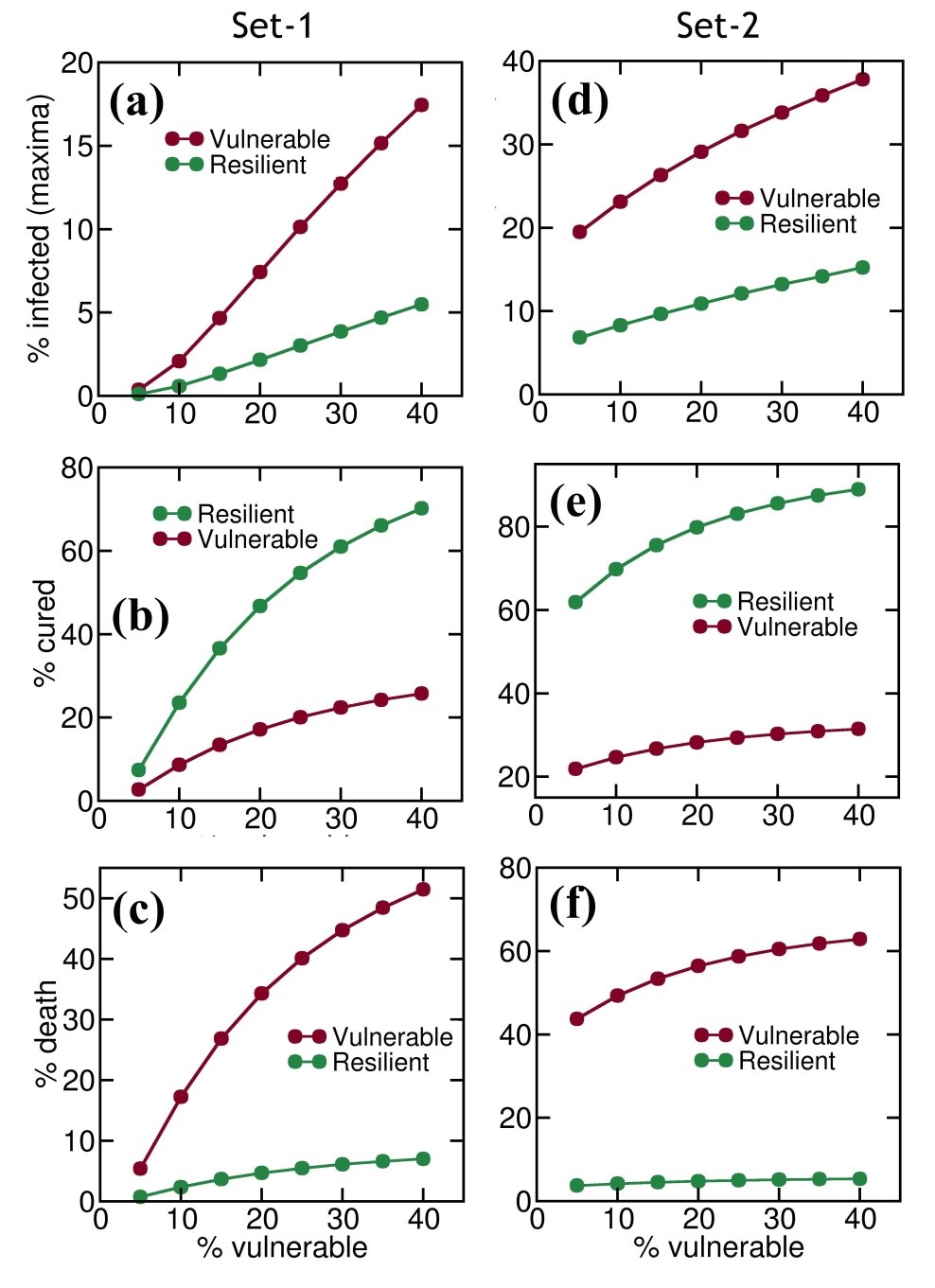}
 \caption{The effect of the change in the initial percentage of the vulnerable population on the relative infection and recovery for the sub-categories, namely, vulnerables and resilients. Plots show the dependence of infection peak, percentage cured and dead population for vulnerable (maroon) and resilient (green) population with the initial fraction of vulnerable population as obtained from the solution of the modified SIR model described in Eqs. \ref{eq4} - \ref{eq11}. For figures (a)-(c) set-1 and for (d)-(f) set-2. The quantities show a non-linear dependence and enhanced fatality for the vurnerables.}
 \label{fig5}
\end{figure}

Next we vary the \% of initial vulnerable population from 5\% to 40\% and obtain the \% of highest active cases (that is the maxima in the temporal variation of $I_V(t)$ or $I_R(t)$), \% of cured population and \% of death. The range is chosen in order to represent different regions/countries. For example, in India only $\sim8\%$ of the entire population is above 60 years whereas, in countries like Italy and Germany the number is over 20\%.

We obtain Fig. \ref{fig5}a - \ref{fig5}c for set-1 and Fig. \ref{fig5}d - \ref{fig5}f for set-2. In both the cases the variation of the infected peak maxima with \% vulnerable shows nearly linear increase with a higher slope for the vulnerables (Fig. \ref{fig5}a and \ref{fig5}d). Interestingly the \% cured (Fig. \ref{fig5}b and \ref{fig5}e) and \% dead (Fig. \ref{fig5}c and \ref{fig5}f) shows a nonlinear dependence on \% vulnerable. It clearly shows that the damage is huge to the vulnerable population when the \% of vulnerables increases.

\begin{figure}[H]
 \centering
 \includegraphics[width=3.2in,keepaspectratio=true]{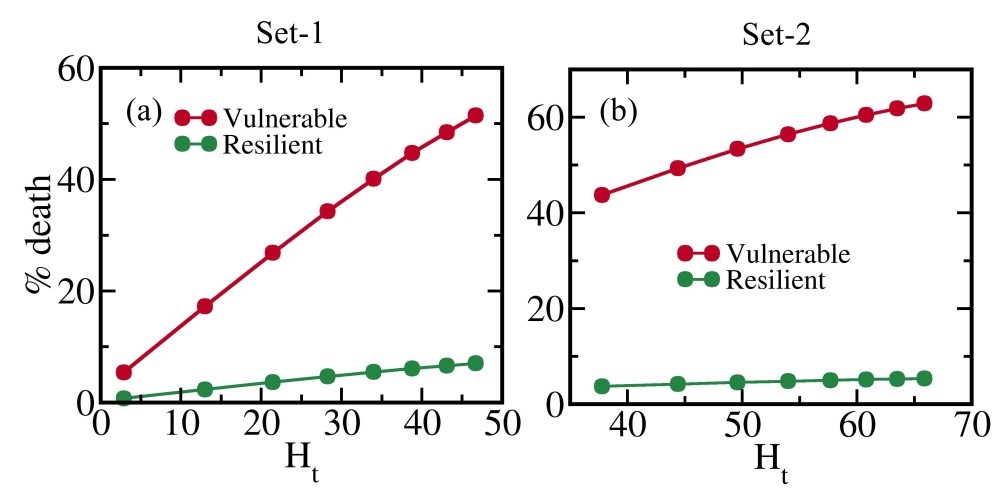}
 \caption{Percentage outcome of different herd immunity thresholds ($H_t$) on the vulberable and resilient population. Plot of percentage deaths against $H_t$ calculated from Eq. \ref{eq2} for (a) set-1, and (b) set-2. In both the two cases the dependence is linear with substantially more damage to the vulnerable population. The values on the Y axes are individually normalised.}
 \label{fig6}
\end{figure}

We plot (Fig. \ref{fig6}) the percentage of deaths for both the subcategories against the herd immunity threshold for a given Vul:Res composition (1:4). This is to show the increasing damage with respect to Ht. We find that the trend is linear for both the sets of parameters and the relative fatality is substantially higher for the vulnerables.

\subsection{(Stochastic Cellular Automata Simulations}
\label{sec3B}
\subsubsection{Dependence on the initial population distribution}

\begin{figure}[H]
 \centering
 \includegraphics[width=2.5in,keepaspectratio=true]{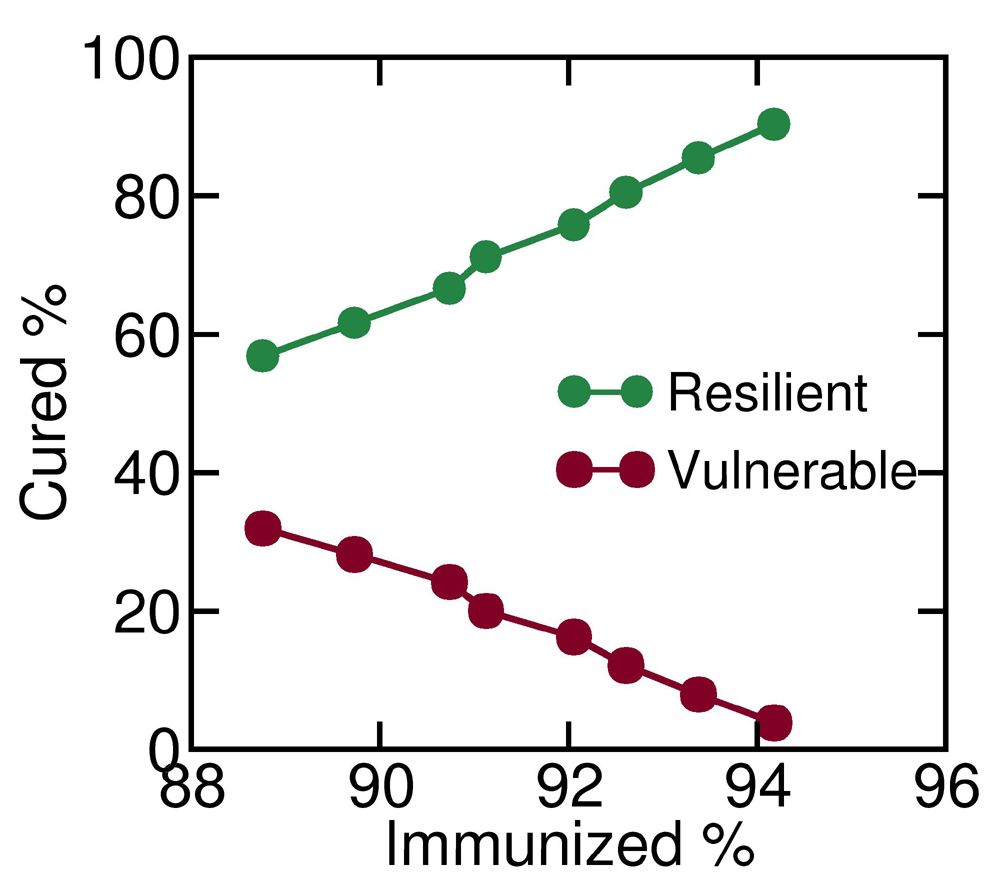}
 \caption{Percentage of cured resilient and vulnerables in the population on the course of attaining herd immunity. The percentage of cured individuals is shown as a function of the percentage of total population immunized after getting infected. This is obtained by averaging over 100 CA simulations. Green shows the percentage of death for the resilient fraction of the society and maroon denotes the same for the vulnerable people.}
 \label{fig7}
\end{figure}

Here, we keep the probability of transmission of disease time-independent and equal for both resilients and vulnerables. We change the initial fraction of the vulnerable section of the total population from 5\% to 40\%. In Fig. \ref{fig7} we plot the \% of cured individuals (resilients and vulnerables) against \% of total immunization when the temporal progression of the population reaches saturation. As discussed earlier, herd immunity is obtained when a major section of the population becomes immune, post infection. However, apart from gaining immunity, this process involves the death of many infected individuals according to their survival probability. The probability of recovery of the resilients is higher than that of the vulnerables. Here, these two probabilities are taken as 0.95 and 0.8 respectively.\cite{Verity2020,Ruan2020}

In Fig. \ref{fig7} the abscissa is the percentage of the total population that becomes immune after recovering from the infection. The ordinate quantifies the percentage of cured resilients and vulnerables with respect to the total initial population. \textit{With increase in the immunity attained in the society, a significant decrease in the percentage of cured vulnerable individuals is observed}. This implies that higher the percentage of immunization in the total population, greater is the probability of death of the vulnerable section. Hence herd immunity results in the death of a major fraction of the vulnerable population. This stratum of the society includes mainly the old people (age greater than years) and people with serious health conditions or comorbidity.\cite{Fang2020,Yang2020} The geographical regions with demographic distributions having higher fraction of the people of age above $\sim$60 years are among the worst affected. For example, Italy suffered the loss of many aged people as a result of the COVID-19 pandemic.\cite{Livingston2020,Onder2020}

\begin{figure}[H]
 \centering
 \includegraphics[width=3.4in,keepaspectratio=true]{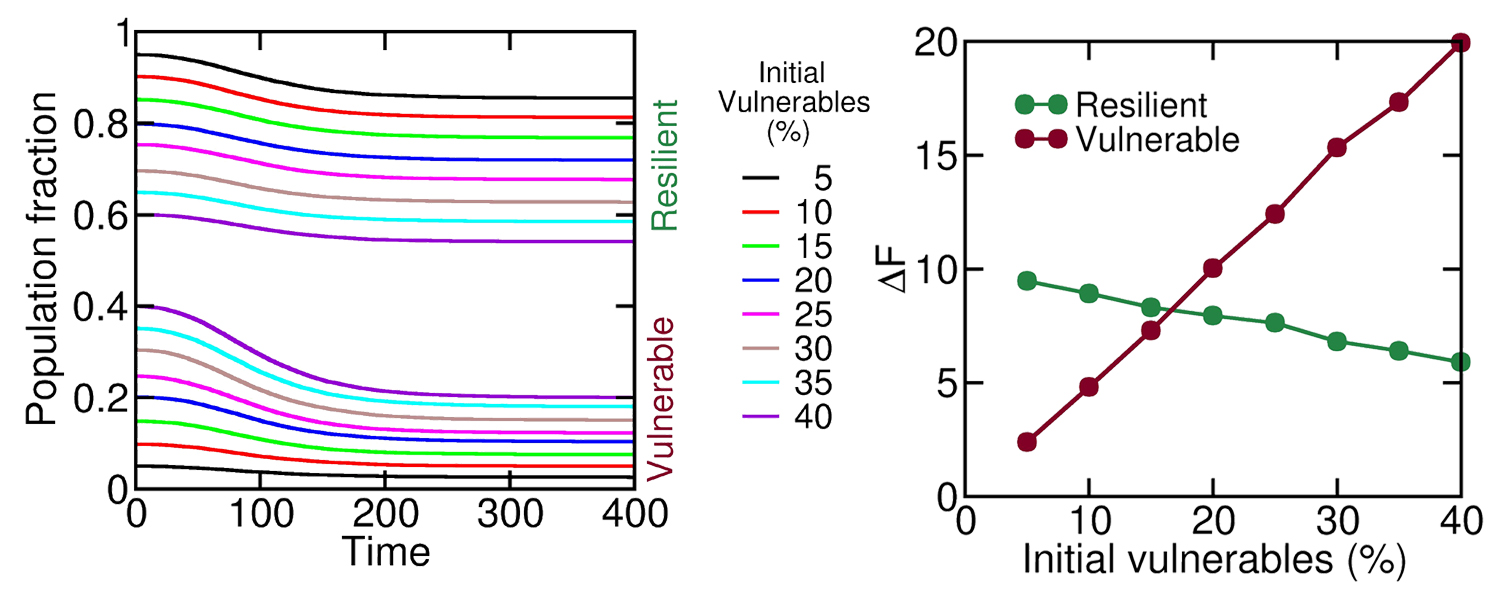}
 \caption{(a) Population dynamics represented as the temporal evolution of the fraction of resilient and vulnerable sections of the population are shown with varying initial distribution of resilients and vulnerables. The colour bar on the right hand side shows the initial \% of vulnerables in the total population. (b) The absolute decrease in the resilient (green) and vulnerable (maroon) fractions of the total population as functions of the initial percentage of vulnerables.}
 \label{fig8}
\end{figure}

In Fig. \ref{fig8}a , we show the time evolution of the fraction of vulnerables and resilients in the total population for different \% of initial number of vulnerables. The fractions are calculated with respect to the total initial population. We see that with increase in the initial \% of vulnerables, the number of resilients dying show a slight decrease, whereas the number of dead vulnerables increases significantly. This observation is clarified in Fig. \ref{fig8}b. Here we plot the absolute change in the fraction of resilients and vulnerables as functions of the initial \% of vulnerables. Both show linear dependence. The gradient (slope) is negative for resilients and positive for vulnerables. However, we find that the absolute value of the slope for the latter is $\sim$5 times higher than that of the former. This denotes that countries with higher population of elderly and vulnerable people in the society incur a greater loss in the number of vulnerable individuals.

\subsubsection{Dependence on the probability of recovery}

Now, we keep the initial population distribution fixed at 20\% vulnerable and 80\% resilient individuals. We change the probability of recovery of these two categories ($P_R^{Vul}$ and $P_R^{Res}$) with the constraint $P_R^{Vul} \leq P_R^{Res}$. Accordingly, we change these two probabilities from 0.6 to 0.8 and 0.8 to 0.95 respectively. We choose these values according to reported case fatality ratios for the SARS-CoV-2 pandemic.\cite{Verity2020, Ruan2020}

\begin{figure}[ht]
 \centering
 \includegraphics[width=3in,keepaspectratio=true]{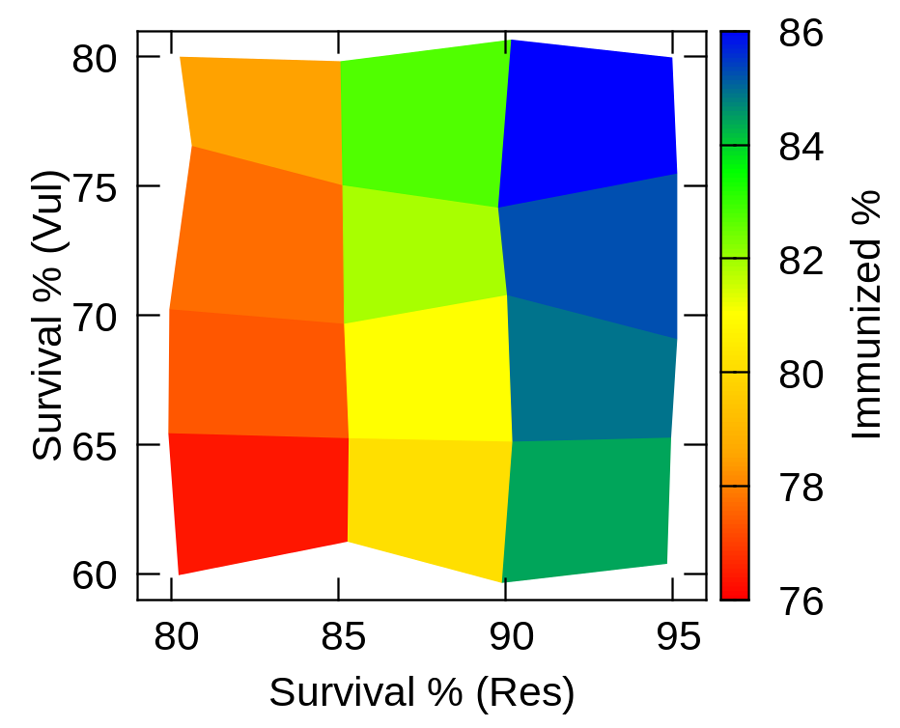}
 \caption{Interdependence of different fractions of the population as the immunity evolves. Percentage of immunized (colour coded) represented as a function of the percentage of survival for vulnerables and resilients. The proportions are with respect to the total initial population. The primary variables are the probabilities of recovery of the vulnerables and the resilients. The results are obtained after averaging over 100 simulations.}
 \label{fig9}
\end{figure}

For every pair of $P_R^{Vul}$ and $P_R^{Res}$ we get a value of percentage of vulnerables and resilients who survive and a fraction of the population that gets immunized. In Figure 9 we plot the survival \% of vulnerables and resilients in the two perpendicular axes and represent the \% immunized as colour codes according to the colour gradation bar on the right hand side. In this contour representation, red denotes low immunity and blue denotes higher immunity.

The survival \% of the vulnerables is lower than that of the resilients. The percentage of immunized population is higher (blue) for maximum survival of the resilients as compared to that of the vulnerables. This means that to attain higher immunity in the population, greater number of old and vulnerable people suffer death as compared to resilients. Hence, attainment of herd immunity comes with the cost of a higher mortality of the vulnerable section of the society.

\section{Summary and Conclusion}

Any epidemic is a dynamic process where time dependence plays a crucial role in the control of the spread and the damage, that is, the outcome. COVID-19 is a pandemic which is currently under intense scrutiny by all and sundry, and many aspects are yet to be understood. Every move by the government, and the population in general, is of crucial importance. Each pandemic comes with unique characteristics that deserve special treatments, not just medical and clinical but also sociological. In each such epidemic, immunity plays a critical role. Spanish Flu mainly attacked the age group between 20 and 30 years of age. This is the age group with maximum immunity. In the case of COVID-19, again we face the sad reality that certain section of the society is substantially more vulnerable than other sections. The vulnerable section consists of age groups which are above 60-65 years of age, and people with comorbidity. There is yet to further classification, although it is conceivable that as we understand the disease better and more precisely, better perception of danger would emerge.

An epidemic often starts by a process of nucleation which is an important phenomenon often studied in physics and chemistry. The process of nucleation is initiated by a sudden appearance of a group of infected individuals in a region. This may be triggered a laboratory accident, or infection from eating wild animals like bats, pangolin etc. or by arrival of infected tourists and so on. The process may be dependent on the nature of the geography and demography of the country or region. The initial period of the process is often slow. After the initial nucleation, the disease spreads by a diffusion process into the susceptible population. Hence, it is a percolation with a temporal evolution.

In order to address the issue of vulnerability of the population and the outcome with the progression of the epidemic, we carry out a theoretical analysis with the objective to analyze the consequences of aiming for herd immunity without vaccine, or a good drug, in the context of the present COVID-19 pandemic. We develop and solve a modified SIR model numerically and by employing cellular automata simulations. We particularly probed the following question: what is dependence of mortality on the rate of herd immunity?

One of the key results of the present study is the dependence of the percentage survival on the rate of attainment of the immunity threshold. We find that a late attainment of the immunity saturation leads to relatively lesser fatality. We show that approximately 50-60\% of the vulnerables might lose their lives in order to attain ~70\% total immunized population. On the contrary the mortality of the resilient fraction of population is relatively low, may be just about 10\%. We find a non-linear trend in the dependence of the cured and dead population on the initial population of the vulnerables. This is because as the number of vulnerables increases, the immunity by infection from a larger fraction of population which cannot protect the vulnerables unless deliberate efforts are made that requires intervention.

While we discuss herd immunity by infection in this work, the other, more sustainable option is herd immunity by vaccination. For example, diseases like small pox, polio etc. have been wiped off the face of earth by vaccination. This is particularly crucial for diseases with high mortality rates. However, for any novel disease, preparation of a vaccine can take years. In case of the present COVID-19 pandemic, for instance, extensive research is going on globally in search of a vaccine.\cite{Chen2020} However, no promising result has been obtained in almost five months and researchers believe it may take more than a year to prepare the vaccine. 

\begin{acknowledgments}
We thank Prof. Sarika Bhattacharyya (NCL, Pune) and Prof. Suman Chakrabarty (SNCNCBS, Kolkata) for several fruitful discussions and comments. The authors thank the Department of Science and Technology (DST), India for financial support. BB thanks sir J. C. Bose fellowship for partial financial support. S. Mo. thanks Universities Grants Commission (UGC), India for research fellowship. S. Mu. thanks DST-INSPIRE programme for providing research fellowship.
\end{acknowledgments}

\bibliography{ref}

\begin{thebibliography}{41}%
\makeatletter
\providecommand \@ifxundefined [1]{%
 \@ifx{#1\undefined}
}%
\providecommand \@ifnum [1]{%
 \ifnum #1\expandafter \@firstoftwo
 \else \expandafter \@secondoftwo
 \fi
}%
\providecommand \@ifx [1]{%
 \ifx #1\expandafter \@firstoftwo
 \else \expandafter \@secondoftwo
 \fi
}%
\providecommand \natexlab [1]{#1}%
\providecommand \enquote  [1]{``#1''}%
\providecommand \bibnamefont  [1]{#1}%
\providecommand \bibfnamefont [1]{#1}%
\providecommand \citenamefont [1]{#1}%
\providecommand \href@noop [0]{\@secondoftwo}%
\providecommand \href [0]{\begingroup \@sanitize@url \@href}%
\providecommand \@href[1]{\@@startlink{#1}\@@href}%
\providecommand \@@href[1]{\endgroup#1\@@endlink}%
\providecommand \@sanitize@url [0]{\catcode `\\12\catcode `\$12\catcode
  `\&12\catcode `\#12\catcode `\^12\catcode `\_12\catcode `\%12\relax}%
\providecommand \@@startlink[1]{}%
\providecommand \@@endlink[0]{}%
\providecommand \url  [0]{\begingroup\@sanitize@url \@url }%
\providecommand \@url [1]{\endgroup\@href {#1}{\urlprefix }}%
\providecommand \urlprefix  [0]{URL }%
\providecommand \Eprint [0]{\href }%
\providecommand \doibase [0]{http://dx.doi.org/}%
\providecommand \selectlanguage [0]{\@gobble}%
\providecommand \bibinfo  [0]{\@secondoftwo}%
\providecommand \bibfield  [0]{\@secondoftwo}%
\providecommand \translation [1]{[#1]}%
\providecommand \BibitemOpen [0]{}%
\providecommand \bibitemStop [0]{}%
\providecommand \bibitemNoStop [0]{.\EOS\space}%
\providecommand \EOS [0]{\spacefactor3000\relax}%
\providecommand \BibitemShut  [1]{\csname bibitem#1\endcsname}%
\let\auto@bib@innerbib\@empty
\bibitem [{\citenamefont {Chen}\ \emph {et~al.}(2020)\citenamefont {Chen},
  \citenamefont {Strych}, \citenamefont {Hotez},\ and\ \citenamefont
  {Bottazzi}}]{Chen2020}%
  \BibitemOpen
  \bibfield  {author} {\bibinfo {author} {\bibfnamefont {W.-H.}\ \bibnamefont
  {Chen}}, \bibinfo {author} {\bibfnamefont {U.}~\bibnamefont {Strych}},
  \bibinfo {author} {\bibfnamefont {P.~J.}\ \bibnamefont {Hotez}}, \ and\
  \bibinfo {author} {\bibfnamefont {M.~E.}\ \bibnamefont {Bottazzi}},\
  }\href@noop {} {\bibfield  {journal} {\bibinfo  {journal} {Current tropical
  medicine reports}\ ,\ \bibinfo {pages} {1}} (\bibinfo {year}
  {2020})}\BibitemShut {NoStop}%
\bibitem [{\citenamefont {Prem}\ \emph {et~al.}(2020)\citenamefont {Prem},
  \citenamefont {Liu}, \citenamefont {Russell}, \citenamefont {Kucharski},
  \citenamefont {Eggo}, \citenamefont {Davies}, \citenamefont {Flasche},
  \citenamefont {Clifford}, \citenamefont {Pearson}, \citenamefont {Munday}
  \emph {et~al.}}]{Prem2020}%
  \BibitemOpen
  \bibfield  {author} {\bibinfo {author} {\bibfnamefont {K.}~\bibnamefont
  {Prem}}, \bibinfo {author} {\bibfnamefont {Y.}~\bibnamefont {Liu}}, \bibinfo
  {author} {\bibfnamefont {T.~W.}\ \bibnamefont {Russell}}, \bibinfo {author}
  {\bibfnamefont {A.~J.}\ \bibnamefont {Kucharski}}, \bibinfo {author}
  {\bibfnamefont {R.~M.}\ \bibnamefont {Eggo}}, \bibinfo {author}
  {\bibfnamefont {N.}~\bibnamefont {Davies}}, \bibinfo {author} {\bibfnamefont
  {S.}~\bibnamefont {Flasche}}, \bibinfo {author} {\bibfnamefont
  {S.}~\bibnamefont {Clifford}}, \bibinfo {author} {\bibfnamefont {C.~A.}\
  \bibnamefont {Pearson}}, \bibinfo {author} {\bibfnamefont {J.~D.}\
  \bibnamefont {Munday}},  \emph {et~al.},\ }\href@noop {} {\bibfield
  {journal} {\bibinfo  {journal} {The Lancet Public Health}\ } (\bibinfo {year}
  {2020})}\BibitemShut {NoStop}%
\bibitem [{\citenamefont {Shen}\ \emph {et~al.}(2020)\citenamefont {Shen},
  \citenamefont {Peng}, \citenamefont {Guo}, \citenamefont {Xiao},\ and\
  \citenamefont {Zhang}}]{Shen2020}%
  \BibitemOpen
  \bibfield  {author} {\bibinfo {author} {\bibfnamefont {M.}~\bibnamefont
  {Shen}}, \bibinfo {author} {\bibfnamefont {Z.}~\bibnamefont {Peng}}, \bibinfo
  {author} {\bibfnamefont {Y.}~\bibnamefont {Guo}}, \bibinfo {author}
  {\bibfnamefont {Y.}~\bibnamefont {Xiao}}, \ and\ \bibinfo {author}
  {\bibfnamefont {L.}~\bibnamefont {Zhang}},\ }\href@noop {} {\bibfield
  {journal} {\bibinfo  {journal} {medRxiv}\ } (\bibinfo {year}
  {2020})}\BibitemShut {NoStop}%
\bibitem [{\citenamefont {Kamikubo}\ and\ \citenamefont
  {Takahashi}(2020)}]{Kamikubo2020}%
  \BibitemOpen
  \bibfield  {author} {\bibinfo {author} {\bibfnamefont {Y.}~\bibnamefont
  {Kamikubo}}\ and\ \bibinfo {author} {\bibfnamefont {A.}~\bibnamefont
  {Takahashi}},\ }\href@noop {} {\bibfield  {journal} {\bibinfo  {journal}
  {medRxiv}\ } (\bibinfo {year} {2020})}\BibitemShut {NoStop}%
\bibitem [{\citenamefont {Fine}(1993)}]{Fine1993}%
  \BibitemOpen
  \bibfield  {author} {\bibinfo {author} {\bibfnamefont {P.~E.}\ \bibnamefont
  {Fine}},\ }\href@noop {} {\bibfield  {journal} {\bibinfo  {journal}
  {Epidemiologic reviews}\ }\textbf {\bibinfo {volume} {15}},\ \bibinfo {pages}
  {265} (\bibinfo {year} {1993})}\BibitemShut {NoStop}%
\bibitem [{\citenamefont {Anderson}\ and\ \citenamefont
  {May}(1985)}]{Anderson1985}%
  \BibitemOpen
  \bibfield  {author} {\bibinfo {author} {\bibfnamefont {R.~M.}\ \bibnamefont
  {Anderson}}\ and\ \bibinfo {author} {\bibfnamefont {R.~M.}\ \bibnamefont
  {May}},\ }\href@noop {} {\bibfield  {journal} {\bibinfo  {journal} {Nature}\
  }\textbf {\bibinfo {volume} {318}},\ \bibinfo {pages} {323} (\bibinfo {year}
  {1985})}\BibitemShut {NoStop}%
\bibitem [{\citenamefont {John}\ and\ \citenamefont {Samuel}(2000)}]{John2000}%
  \BibitemOpen
  \bibfield  {author} {\bibinfo {author} {\bibfnamefont {T.~J.}\ \bibnamefont
  {John}}\ and\ \bibinfo {author} {\bibfnamefont {R.}~\bibnamefont {Samuel}},\
  }\href@noop {} {\bibfield  {journal} {\bibinfo  {journal} {European journal
  of epidemiology}\ }\textbf {\bibinfo {volume} {16}},\ \bibinfo {pages} {601}
  (\bibinfo {year} {2000})}\BibitemShut {NoStop}%
\bibitem [{\citenamefont {Georgette}(2009)}]{Georgette2009}%
  \BibitemOpen
  \bibfield  {author} {\bibinfo {author} {\bibfnamefont {N.~T.}\ \bibnamefont
  {Georgette}},\ }\href@noop {} {\bibfield  {journal} {\bibinfo  {journal}
  {PloS one}\ }\textbf {\bibinfo {volume} {4}},\ \bibinfo {pages} {e4168}
  (\bibinfo {year} {2009})}\BibitemShut {NoStop}%
\bibitem [{\citenamefont {McBryde}(2009)}]{McBryde2009}%
  \BibitemOpen
  \bibfield  {author} {\bibinfo {author} {\bibfnamefont {E.~S.}\ \bibnamefont
  {McBryde}},\ }\href@noop {} {\bibfield  {journal} {\bibinfo  {journal}
  {Clinical infectious diseases}\ }\textbf {\bibinfo {volume} {48}},\ \bibinfo
  {pages} {685} (\bibinfo {year} {2009})}\BibitemShut {NoStop}%
\bibitem [{\citenamefont {Wrapp}\ \emph {et~al.}(2020)\citenamefont {Wrapp},
  \citenamefont {Wang}, \citenamefont {Corbett}, \citenamefont {Goldsmith},
  \citenamefont {Hsieh}, \citenamefont {Abiona}, \citenamefont {Graham},\ and\
  \citenamefont {McLellan}}]{Wrapp2020}%
  \BibitemOpen
  \bibfield  {author} {\bibinfo {author} {\bibfnamefont {D.}~\bibnamefont
  {Wrapp}}, \bibinfo {author} {\bibfnamefont {N.}~\bibnamefont {Wang}},
  \bibinfo {author} {\bibfnamefont {K.~S.}\ \bibnamefont {Corbett}}, \bibinfo
  {author} {\bibfnamefont {J.~A.}\ \bibnamefont {Goldsmith}}, \bibinfo {author}
  {\bibfnamefont {C.-L.}\ \bibnamefont {Hsieh}}, \bibinfo {author}
  {\bibfnamefont {O.}~\bibnamefont {Abiona}}, \bibinfo {author} {\bibfnamefont
  {B.~S.}\ \bibnamefont {Graham}}, \ and\ \bibinfo {author} {\bibfnamefont
  {J.~S.}\ \bibnamefont {McLellan}},\ }\href@noop {} {\bibfield  {journal}
  {\bibinfo  {journal} {Science}\ }\textbf {\bibinfo {volume} {367}},\ \bibinfo
  {pages} {1260} (\bibinfo {year} {2020})}\BibitemShut {NoStop}%
\bibitem [{\citenamefont {Singh}\ and\ \citenamefont
  {Adhikari}(2020)}]{Singh2020}%
  \BibitemOpen
  \bibfield  {author} {\bibinfo {author} {\bibfnamefont {R.}~\bibnamefont
  {Singh}}\ and\ \bibinfo {author} {\bibfnamefont {R.}~\bibnamefont
  {Adhikari}},\ }\href@noop {} {\bibfield  {journal} {\bibinfo  {journal}
  {arXiv preprint arXiv:2003.12055}\ } (\bibinfo {year} {2020})}\BibitemShut
  {NoStop}%
\bibitem [{\citenamefont {Mukherjee}\ \emph {et~al.}(2020)\citenamefont
  {Mukherjee}, \citenamefont {Mondal},\ and\ \citenamefont
  {Bagchi}}]{Mukherjee2020}%
  \BibitemOpen
  \bibfield  {author} {\bibinfo {author} {\bibfnamefont {S.}~\bibnamefont
  {Mukherjee}}, \bibinfo {author} {\bibfnamefont {S.}~\bibnamefont {Mondal}}, \
  and\ \bibinfo {author} {\bibfnamefont {B.}~\bibnamefont {Bagchi}},\
  }\href@noop {} {\bibfield  {journal} {\bibinfo  {journal} {arXiv preprint
  arXiv:2004.14787}\ } (\bibinfo {year} {2020})}\BibitemShut {NoStop}%
\bibitem [{\citenamefont {Daley}\ and\ \citenamefont {Gani}(2001)}]{Daley2001}%
  \BibitemOpen
  \bibfield  {author} {\bibinfo {author} {\bibfnamefont {D.~J.}\ \bibnamefont
  {Daley}}\ and\ \bibinfo {author} {\bibfnamefont {J.}~\bibnamefont {Gani}},\
  }\href@noop {} {\emph {\bibinfo {title} {Epidemic modelling: an
  introduction}}},\ Vol.~\bibinfo {volume} {15}\ (\bibinfo  {publisher}
  {Cambridge University Press},\ \bibinfo {year} {2001})\BibitemShut {NoStop}%
\bibitem [{\citenamefont {Kermack}\ and\ \citenamefont
  {McKendrick}(1927)}]{Kermack1927}%
  \BibitemOpen
  \bibfield  {author} {\bibinfo {author} {\bibfnamefont {W.~O.}\ \bibnamefont
  {Kermack}}\ and\ \bibinfo {author} {\bibfnamefont {A.~G.}\ \bibnamefont
  {McKendrick}},\ }\href@noop {} {\bibfield  {journal} {\bibinfo  {journal}
  {Proceedings of the royal society of london. Series A, Containing papers of a
  mathematical and physical character}\ }\textbf {\bibinfo {volume} {115}},\
  \bibinfo {pages} {700} (\bibinfo {year} {1927})}\BibitemShut {NoStop}%
\bibitem [{\citenamefont {Skvortsov}\ \emph {et~al.}(2007)\citenamefont
  {Skvortsov}, \citenamefont {Connell}, \citenamefont {Dawson},\ and\
  \citenamefont {Gailis}}]{Skvortsov2007}%
  \BibitemOpen
  \bibfield  {author} {\bibinfo {author} {\bibfnamefont {A.}~\bibnamefont
  {Skvortsov}}, \bibinfo {author} {\bibfnamefont {R.}~\bibnamefont {Connell}},
  \bibinfo {author} {\bibfnamefont {P.}~\bibnamefont {Dawson}}, \ and\ \bibinfo
  {author} {\bibfnamefont {R.}~\bibnamefont {Gailis}},\ }in\ \href@noop {}
  {\emph {\bibinfo {booktitle} {MODSIM 2007 International Congress on Modelling
  and Simulation. Modelling and Simulation Society of Australia and New
  Zealand}}}\ (\bibinfo {organization} {Citeseer},\ \bibinfo {year} {2007})\
  pp.\ \bibinfo {pages} {657--662}\BibitemShut {NoStop}%
\bibitem [{\citenamefont {Jones}\ \emph {et~al.}(2009)\citenamefont {Jones},
  \citenamefont {Plank},\ and\ \citenamefont {Sleeman}}]{Jones2009}%
  \BibitemOpen
  \bibfield  {author} {\bibinfo {author} {\bibfnamefont {D.~S.}\ \bibnamefont
  {Jones}}, \bibinfo {author} {\bibfnamefont {M.}~\bibnamefont {Plank}}, \ and\
  \bibinfo {author} {\bibfnamefont {B.~D.}\ \bibnamefont {Sleeman}},\
  }\href@noop {} {\emph {\bibinfo {title} {Differential equations and
  mathematical biology}}}\ (\bibinfo  {publisher} {CRC press},\ \bibinfo {year}
  {2009})\BibitemShut {NoStop}%
\bibitem [{\citenamefont {Anderson}\ and\ \citenamefont
  {May}(1979)}]{Anderson1979}%
  \BibitemOpen
  \bibfield  {author} {\bibinfo {author} {\bibfnamefont {R.~M.}\ \bibnamefont
  {Anderson}}\ and\ \bibinfo {author} {\bibfnamefont {R.~M.}\ \bibnamefont
  {May}},\ }\href@noop {} {\bibfield  {journal} {\bibinfo  {journal} {Nature}\
  }\textbf {\bibinfo {volume} {280}},\ \bibinfo {pages} {361} (\bibinfo {year}
  {1979})}\BibitemShut {NoStop}%
\bibitem [{\citenamefont {Dietz}(1993)}]{Dietz1993}%
  \BibitemOpen
  \bibfield  {author} {\bibinfo {author} {\bibfnamefont {K.}~\bibnamefont
  {Dietz}},\ }\href@noop {} {\bibfield  {journal} {\bibinfo  {journal}
  {Statistical methods in medical research}\ }\textbf {\bibinfo {volume} {2}},\
  \bibinfo {pages} {23} (\bibinfo {year} {1993})}\BibitemShut {NoStop}%
\bibitem [{\citenamefont {Diekmann}\ \emph {et~al.}(1995)\citenamefont
  {Diekmann}, \citenamefont {Heesterbeek},\ and\ \citenamefont
  {Metz}}]{Diekmann1995}%
  \BibitemOpen
  \bibfield  {author} {\bibinfo {author} {\bibfnamefont {O.}~\bibnamefont
  {Diekmann}}, \bibinfo {author} {\bibfnamefont {J.~A.~P.}\ \bibnamefont
  {Heesterbeek}}, \ and\ \bibinfo {author} {\bibfnamefont {J.~A.}\ \bibnamefont
  {Metz}},\ }\href@noop {} {\bibfield  {journal} {\bibinfo  {journal}
  {Publications of the Newton Institute}\ }\textbf {\bibinfo {volume} {5}},\
  \bibinfo {pages} {95} (\bibinfo {year} {1995})}\BibitemShut {NoStop}%
\bibitem [{\citenamefont {Zhang}\ \emph {et~al.}(2020)\citenamefont {Zhang},
  \citenamefont {Diao}, \citenamefont {Yu}, \citenamefont {Pei}, \citenamefont
  {Lin},\ and\ \citenamefont {Chen}}]{Zhang2020}%
  \BibitemOpen
  \bibfield  {author} {\bibinfo {author} {\bibfnamefont {S.}~\bibnamefont
  {Zhang}}, \bibinfo {author} {\bibfnamefont {M.}~\bibnamefont {Diao}},
  \bibinfo {author} {\bibfnamefont {W.}~\bibnamefont {Yu}}, \bibinfo {author}
  {\bibfnamefont {L.}~\bibnamefont {Pei}}, \bibinfo {author} {\bibfnamefont
  {Z.}~\bibnamefont {Lin}}, \ and\ \bibinfo {author} {\bibfnamefont
  {D.}~\bibnamefont {Chen}},\ }\href@noop {} {\bibfield  {journal} {\bibinfo
  {journal} {International Journal of Infectious Diseases}\ }\textbf {\bibinfo
  {volume} {93}},\ \bibinfo {pages} {201} (\bibinfo {year} {2020})}\BibitemShut
  {NoStop}%
\bibitem [{\citenamefont {Tang}\ \emph {et~al.}(2020)\citenamefont {Tang},
  \citenamefont {Bragazzi}, \citenamefont {Li}, \citenamefont {Tang},
  \citenamefont {Xiao},\ and\ \citenamefont {Wu}}]{Tang2020}%
  \BibitemOpen
  \bibfield  {author} {\bibinfo {author} {\bibfnamefont {B.}~\bibnamefont
  {Tang}}, \bibinfo {author} {\bibfnamefont {N.~L.}\ \bibnamefont {Bragazzi}},
  \bibinfo {author} {\bibfnamefont {Q.}~\bibnamefont {Li}}, \bibinfo {author}
  {\bibfnamefont {S.}~\bibnamefont {Tang}}, \bibinfo {author} {\bibfnamefont
  {Y.}~\bibnamefont {Xiao}}, \ and\ \bibinfo {author} {\bibfnamefont
  {J.}~\bibnamefont {Wu}},\ }\href@noop {} {\bibfield  {journal} {\bibinfo
  {journal} {Infectious disease modelling}\ }\textbf {\bibinfo {volume} {5}},\
  \bibinfo {pages} {248} (\bibinfo {year} {2020})}\BibitemShut {NoStop}%
\bibitem [{\citenamefont {Yang}\ \emph {et~al.}(2020)\citenamefont {Yang},
  \citenamefont {Zheng}, \citenamefont {Gou}, \citenamefont {Pu}, \citenamefont
  {Chen}, \citenamefont {Guo}, \citenamefont {Ji}, \citenamefont {Wang},
  \citenamefont {Wang},\ and\ \citenamefont {Zhou}}]{Yang2020}%
  \BibitemOpen
  \bibfield  {author} {\bibinfo {author} {\bibfnamefont {J.}~\bibnamefont
  {Yang}}, \bibinfo {author} {\bibfnamefont {Y.}~\bibnamefont {Zheng}},
  \bibinfo {author} {\bibfnamefont {X.}~\bibnamefont {Gou}}, \bibinfo {author}
  {\bibfnamefont {K.}~\bibnamefont {Pu}}, \bibinfo {author} {\bibfnamefont
  {Z.}~\bibnamefont {Chen}}, \bibinfo {author} {\bibfnamefont {Q.}~\bibnamefont
  {Guo}}, \bibinfo {author} {\bibfnamefont {R.}~\bibnamefont {Ji}}, \bibinfo
  {author} {\bibfnamefont {H.}~\bibnamefont {Wang}}, \bibinfo {author}
  {\bibfnamefont {Y.}~\bibnamefont {Wang}}, \ and\ \bibinfo {author}
  {\bibfnamefont {Y.}~\bibnamefont {Zhou}},\ }\href@noop {} {\bibfield
  {journal} {\bibinfo  {journal} {International Journal of Infectious
  Diseases}\ } (\bibinfo {year} {2020})}\BibitemShut {NoStop}%
\bibitem [{\citenamefont {Wangping}\ \emph {et~al.}(2020)\citenamefont
  {Wangping}, \citenamefont {Ke}, \citenamefont {Yang}, \citenamefont {Wenzhe},
  \citenamefont {Shengshu}, \citenamefont {Shanshan}, \citenamefont {Jianwei},
  \citenamefont {Fuyin}, \citenamefont {Penggang}, \citenamefont {Jing} \emph
  {et~al.}}]{Wangping2020}%
  \BibitemOpen
  \bibfield  {author} {\bibinfo {author} {\bibfnamefont {J.}~\bibnamefont
  {Wangping}}, \bibinfo {author} {\bibfnamefont {H.}~\bibnamefont {Ke}},
  \bibinfo {author} {\bibfnamefont {S.}~\bibnamefont {Yang}}, \bibinfo {author}
  {\bibfnamefont {C.}~\bibnamefont {Wenzhe}}, \bibinfo {author} {\bibfnamefont
  {W.}~\bibnamefont {Shengshu}}, \bibinfo {author} {\bibfnamefont
  {Y.}~\bibnamefont {Shanshan}}, \bibinfo {author} {\bibfnamefont
  {W.}~\bibnamefont {Jianwei}}, \bibinfo {author} {\bibfnamefont
  {K.}~\bibnamefont {Fuyin}}, \bibinfo {author} {\bibfnamefont
  {T.}~\bibnamefont {Penggang}}, \bibinfo {author} {\bibfnamefont
  {L.}~\bibnamefont {Jing}},  \emph {et~al.},\ }\href@noop {} {\bibfield
  {journal} {\bibinfo  {journal} {Frontiers in Medicine}\ }\textbf {\bibinfo
  {volume} {7}},\ \bibinfo {pages} {169} (\bibinfo {year} {2020})}\BibitemShut
  {NoStop}%
\bibitem [{\citenamefont {Lin}\ \emph {et~al.}(2020)\citenamefont {Lin},
  \citenamefont {Zhao}, \citenamefont {Gao}, \citenamefont {Lou}, \citenamefont
  {Yang}, \citenamefont {Musa}, \citenamefont {Wang}, \citenamefont {Cai},
  \citenamefont {Wang}, \citenamefont {Yang} \emph {et~al.}}]{Lin2020}%
  \BibitemOpen
  \bibfield  {author} {\bibinfo {author} {\bibfnamefont {Q.}~\bibnamefont
  {Lin}}, \bibinfo {author} {\bibfnamefont {S.}~\bibnamefont {Zhao}}, \bibinfo
  {author} {\bibfnamefont {D.}~\bibnamefont {Gao}}, \bibinfo {author}
  {\bibfnamefont {Y.}~\bibnamefont {Lou}}, \bibinfo {author} {\bibfnamefont
  {S.}~\bibnamefont {Yang}}, \bibinfo {author} {\bibfnamefont {S.~S.}\
  \bibnamefont {Musa}}, \bibinfo {author} {\bibfnamefont {M.~H.}\ \bibnamefont
  {Wang}}, \bibinfo {author} {\bibfnamefont {Y.}~\bibnamefont {Cai}}, \bibinfo
  {author} {\bibfnamefont {W.}~\bibnamefont {Wang}}, \bibinfo {author}
  {\bibfnamefont {L.}~\bibnamefont {Yang}},  \emph {et~al.},\ }\href@noop {}
  {\bibfield  {journal} {\bibinfo  {journal} {International journal of
  infectious diseases}\ } (\bibinfo {year} {2020})}\BibitemShut {NoStop}%
\bibitem [{\citenamefont {Hollingsworth}\ \emph {et~al.}(2004)\citenamefont
  {Hollingsworth}, \citenamefont {Seybold}, \citenamefont {Kier},\ and\
  \citenamefont {Cheng}}]{Hollingsworth2004}%
  \BibitemOpen
  \bibfield  {author} {\bibinfo {author} {\bibfnamefont {C.~A.}\ \bibnamefont
  {Hollingsworth}}, \bibinfo {author} {\bibfnamefont {P.~G.}\ \bibnamefont
  {Seybold}}, \bibinfo {author} {\bibfnamefont {L.~B.}\ \bibnamefont {Kier}}, \
  and\ \bibinfo {author} {\bibfnamefont {C.-K.}\ \bibnamefont {Cheng}},\
  }\href@noop {} {\bibfield  {journal} {\bibinfo  {journal} {International
  journal of chemical kinetics}\ }\textbf {\bibinfo {volume} {36}},\ \bibinfo
  {pages} {230} (\bibinfo {year} {2004})}\BibitemShut {NoStop}%
\bibitem [{\citenamefont {Seybold}\ \emph {et~al.}(1998)\citenamefont
  {Seybold}, \citenamefont {Kier},\ and\ \citenamefont {Cheng}}]{Seybold1998}%
  \BibitemOpen
  \bibfield  {author} {\bibinfo {author} {\bibfnamefont {P.~G.}\ \bibnamefont
  {Seybold}}, \bibinfo {author} {\bibfnamefont {L.~B.}\ \bibnamefont {Kier}}, \
  and\ \bibinfo {author} {\bibfnamefont {C.-K.}\ \bibnamefont {Cheng}},\
  }\href@noop {} {\bibfield  {journal} {\bibinfo  {journal} {The Journal of
  Physical Chemistry A}\ }\textbf {\bibinfo {volume} {102}},\ \bibinfo {pages}
  {886} (\bibinfo {year} {1998})}\BibitemShut {NoStop}%
\bibitem [{\citenamefont {Wolfram}(1983)}]{Wolfram1983}%
  \BibitemOpen
  \bibfield  {author} {\bibinfo {author} {\bibfnamefont {S.}~\bibnamefont
  {Wolfram}},\ }\href@noop {} {\bibfield  {journal} {\bibinfo  {journal}
  {Reviews of modern physics}\ }\textbf {\bibinfo {volume} {55}},\ \bibinfo
  {pages} {601} (\bibinfo {year} {1983})}\BibitemShut {NoStop}%
\bibitem [{\citenamefont {Bartolozzi}\ and\ \citenamefont
  {Thomas}(2004)}]{Bartolozzi2004}%
  \BibitemOpen
  \bibfield  {author} {\bibinfo {author} {\bibfnamefont {M.}~\bibnamefont
  {Bartolozzi}}\ and\ \bibinfo {author} {\bibfnamefont {A.~W.}\ \bibnamefont
  {Thomas}},\ }\href@noop {} {\bibfield  {journal} {\bibinfo  {journal}
  {Physical review E}\ }\textbf {\bibinfo {volume} {69}},\ \bibinfo {pages}
  {046112} (\bibinfo {year} {2004})}\BibitemShut {NoStop}%
\bibitem [{\citenamefont {Soares-Filho}\ \emph {et~al.}(2002)\citenamefont
  {Soares-Filho}, \citenamefont {Cerqueira},\ and\ \citenamefont
  {Pennachin}}]{Soares-Filho2002}%
  \BibitemOpen
  \bibfield  {author} {\bibinfo {author} {\bibfnamefont {B.~S.}\ \bibnamefont
  {Soares-Filho}}, \bibinfo {author} {\bibfnamefont {G.~C.}\ \bibnamefont
  {Cerqueira}}, \ and\ \bibinfo {author} {\bibfnamefont {C.~L.}\ \bibnamefont
  {Pennachin}},\ }\href@noop {} {\bibfield  {journal} {\bibinfo  {journal}
  {Ecological modelling}\ }\textbf {\bibinfo {volume} {154}},\ \bibinfo {pages}
  {217} (\bibinfo {year} {2002})}\BibitemShut {NoStop}%
\bibitem [{\citenamefont {Goltsev}\ \emph {et~al.}(2010)\citenamefont
  {Goltsev}, \citenamefont {De~Abreu}, \citenamefont {Dorogovtsev},\ and\
  \citenamefont {Mendes}}]{Goltsev2010}%
  \BibitemOpen
  \bibfield  {author} {\bibinfo {author} {\bibfnamefont {A.}~\bibnamefont
  {Goltsev}}, \bibinfo {author} {\bibfnamefont {F.}~\bibnamefont {De~Abreu}},
  \bibinfo {author} {\bibfnamefont {S.}~\bibnamefont {Dorogovtsev}}, \ and\
  \bibinfo {author} {\bibfnamefont {J.}~\bibnamefont {Mendes}},\ }\href@noop {}
  {\bibfield  {journal} {\bibinfo  {journal} {Physical Review E}\ }\textbf
  {\bibinfo {volume} {81}},\ \bibinfo {pages} {061921} (\bibinfo {year}
  {2010})}\BibitemShut {NoStop}%
\bibitem [{\citenamefont {Almeida}\ and\ \citenamefont
  {Macau}(2011)}]{Almeida2011}%
  \BibitemOpen
  \bibfield  {author} {\bibinfo {author} {\bibfnamefont {R.~M.}\ \bibnamefont
  {Almeida}}\ and\ \bibinfo {author} {\bibfnamefont {E.~E.}\ \bibnamefont
  {Macau}},\ }in\ \href@noop {} {\emph {\bibinfo {booktitle} {Journal of
  Physics: Conference Series}}},\ Vol.\ \bibinfo {volume} {285}\ (\bibinfo
  {organization} {IOP Publishing},\ \bibinfo {year} {2011})\ p.\ \bibinfo
  {pages} {012038}\BibitemShut {NoStop}%
\bibitem [{\citenamefont {Fu}\ and\ \citenamefont {Milne}(2003)}]{Fu2003}%
  \BibitemOpen
  \bibfield  {author} {\bibinfo {author} {\bibfnamefont {S.}~\bibnamefont
  {Fu}}\ and\ \bibinfo {author} {\bibfnamefont {G.}~\bibnamefont {Milne}},\
  }in\ \href@noop {} {\emph {\bibinfo {booktitle} {Proc. of the Australian
  Conference on Artificial Life}}}\ (\bibinfo {year} {2003})\BibitemShut
  {NoStop}%
\bibitem [{\citenamefont {White}\ \emph {et~al.}(2007)\citenamefont {White},
  \citenamefont {Del~Rey},\ and\ \citenamefont {S{\'a}nchez}}]{White2007}%
  \BibitemOpen
  \bibfield  {author} {\bibinfo {author} {\bibfnamefont {S.~H.}\ \bibnamefont
  {White}}, \bibinfo {author} {\bibfnamefont {A.~M.}\ \bibnamefont {Del~Rey}},
  \ and\ \bibinfo {author} {\bibfnamefont {G.~R.}\ \bibnamefont
  {S{\'a}nchez}},\ }\href@noop {} {\bibfield  {journal} {\bibinfo  {journal}
  {Applied Mathematics and Computation}\ }\textbf {\bibinfo {volume} {186}},\
  \bibinfo {pages} {193} (\bibinfo {year} {2007})}\BibitemShut {NoStop}%
\bibitem [{\citenamefont {Sirakoulis}\ \emph {et~al.}(2000)\citenamefont
  {Sirakoulis}, \citenamefont {Karafyllidis},\ and\ \citenamefont
  {Thanailakis}}]{Sirakoulis2000}%
  \BibitemOpen
  \bibfield  {author} {\bibinfo {author} {\bibfnamefont {G.~C.}\ \bibnamefont
  {Sirakoulis}}, \bibinfo {author} {\bibfnamefont {I.}~\bibnamefont
  {Karafyllidis}}, \ and\ \bibinfo {author} {\bibfnamefont {A.}~\bibnamefont
  {Thanailakis}},\ }\href@noop {} {\bibfield  {journal} {\bibinfo  {journal}
  {Ecological Modelling}\ }\textbf {\bibinfo {volume} {133}},\ \bibinfo {pages}
  {209} (\bibinfo {year} {2000})}\BibitemShut {NoStop}%
\bibitem [{\citenamefont {Remuzzi}\ and\ \citenamefont
  {Remuzzi}(2020)}]{Remuzzi2020}%
  \BibitemOpen
  \bibfield  {author} {\bibinfo {author} {\bibfnamefont {A.}~\bibnamefont
  {Remuzzi}}\ and\ \bibinfo {author} {\bibfnamefont {G.}~\bibnamefont
  {Remuzzi}},\ }\href@noop {} {\bibfield  {journal} {\bibinfo  {journal} {The
  Lancet}\ } (\bibinfo {year} {2020})}\BibitemShut {NoStop}%
\bibitem [{\citenamefont {Ruan}(2020)}]{Ruan2020}%
  \BibitemOpen
  \bibfield  {author} {\bibinfo {author} {\bibfnamefont {S.}~\bibnamefont
  {Ruan}},\ }\href@noop {} {\bibfield  {journal} {\bibinfo  {journal} {The
  Lancet Infectious Diseases}\ } (\bibinfo {year} {2020})}\BibitemShut
  {NoStop}%
\bibitem [{\citenamefont {Wu}\ \emph {et~al.}(2020)\citenamefont {Wu},
  \citenamefont {Leung}, \citenamefont {Bushman}, \citenamefont {Kishore},
  \citenamefont {Niehus}, \citenamefont {de~Salazar}, \citenamefont {Cowling},
  \citenamefont {Lipsitch},\ and\ \citenamefont {Leung}}]{Wu2020}%
  \BibitemOpen
  \bibfield  {author} {\bibinfo {author} {\bibfnamefont {J.~T.}\ \bibnamefont
  {Wu}}, \bibinfo {author} {\bibfnamefont {K.}~\bibnamefont {Leung}}, \bibinfo
  {author} {\bibfnamefont {M.}~\bibnamefont {Bushman}}, \bibinfo {author}
  {\bibfnamefont {N.}~\bibnamefont {Kishore}}, \bibinfo {author} {\bibfnamefont
  {R.}~\bibnamefont {Niehus}}, \bibinfo {author} {\bibfnamefont {P.~M.}\
  \bibnamefont {de~Salazar}}, \bibinfo {author} {\bibfnamefont {B.~J.}\
  \bibnamefont {Cowling}}, \bibinfo {author} {\bibfnamefont {M.}~\bibnamefont
  {Lipsitch}}, \ and\ \bibinfo {author} {\bibfnamefont {G.~M.}\ \bibnamefont
  {Leung}},\ }\href@noop {} {\bibfield  {journal} {\bibinfo  {journal} {Nature
  Medicine}\ ,\ \bibinfo {pages} {1}} (\bibinfo {year} {2020})}\BibitemShut
  {NoStop}%
\bibitem [{\citenamefont {Verity}\ \emph {et~al.}(2020)\citenamefont {Verity},
  \citenamefont {Okell}, \citenamefont {Dorigatti}, \citenamefont {Winskill},
  \citenamefont {Whittaker}, \citenamefont {Imai}, \citenamefont
  {Cuomo-Dannenburg}, \citenamefont {Thompson}, \citenamefont {Walker},
  \citenamefont {Fu} \emph {et~al.}}]{Verity2020}%
  \BibitemOpen
  \bibfield  {author} {\bibinfo {author} {\bibfnamefont {R.}~\bibnamefont
  {Verity}}, \bibinfo {author} {\bibfnamefont {L.~C.}\ \bibnamefont {Okell}},
  \bibinfo {author} {\bibfnamefont {I.}~\bibnamefont {Dorigatti}}, \bibinfo
  {author} {\bibfnamefont {P.}~\bibnamefont {Winskill}}, \bibinfo {author}
  {\bibfnamefont {C.}~\bibnamefont {Whittaker}}, \bibinfo {author}
  {\bibfnamefont {N.}~\bibnamefont {Imai}}, \bibinfo {author} {\bibfnamefont
  {G.}~\bibnamefont {Cuomo-Dannenburg}}, \bibinfo {author} {\bibfnamefont
  {H.}~\bibnamefont {Thompson}}, \bibinfo {author} {\bibfnamefont {P.~G.}\
  \bibnamefont {Walker}}, \bibinfo {author} {\bibfnamefont {H.}~\bibnamefont
  {Fu}},  \emph {et~al.},\ }\href@noop {} {\bibfield  {journal} {\bibinfo
  {journal} {The Lancet Infectious Diseases}\ } (\bibinfo {year}
  {2020})}\BibitemShut {NoStop}%
\bibitem [{\citenamefont {Fang}\ \emph {et~al.}(2020)\citenamefont {Fang},
  \citenamefont {Karakiulakis},\ and\ \citenamefont {Roth}}]{Fang2020}%
  \BibitemOpen
  \bibfield  {author} {\bibinfo {author} {\bibfnamefont {L.}~\bibnamefont
  {Fang}}, \bibinfo {author} {\bibfnamefont {G.}~\bibnamefont {Karakiulakis}},
  \ and\ \bibinfo {author} {\bibfnamefont {M.}~\bibnamefont {Roth}},\
  }\href@noop {} {\bibfield  {journal} {\bibinfo  {journal} {The Lancet.
  Respiratory Medicine}\ }\textbf {\bibinfo {volume} {8}},\ \bibinfo {pages}
  {e21} (\bibinfo {year} {2020})}\BibitemShut {NoStop}%
\bibitem [{\citenamefont {Livingston}\ and\ \citenamefont
  {Bucher}(2020)}]{Livingston2020}%
  \BibitemOpen
  \bibfield  {author} {\bibinfo {author} {\bibfnamefont {E.}~\bibnamefont
  {Livingston}}\ and\ \bibinfo {author} {\bibfnamefont {K.}~\bibnamefont
  {Bucher}},\ }\href@noop {} {\bibfield  {journal} {\bibinfo  {journal} {Jama}\
  }\textbf {\bibinfo {volume} {323}},\ \bibinfo {pages} {1335} (\bibinfo {year}
  {2020})}\BibitemShut {NoStop}%
\bibitem [{\citenamefont {Onder}\ \emph {et~al.}(2020)\citenamefont {Onder},
  \citenamefont {Rezza},\ and\ \citenamefont {Brusaferro}}]{Onder2020}%
  \BibitemOpen
  \bibfield  {author} {\bibinfo {author} {\bibfnamefont {G.}~\bibnamefont
  {Onder}}, \bibinfo {author} {\bibfnamefont {G.}~\bibnamefont {Rezza}}, \ and\
  \bibinfo {author} {\bibfnamefont {S.}~\bibnamefont {Brusaferro}},\
  }\href@noop {} {\bibfield  {journal} {\bibinfo  {journal} {Jama}\ } (\bibinfo
  {year} {2020})}\BibitemShut {NoStop}%
\end{thebibliography}%


%
\end{document}